\documentclass[manyauthors,nocleardouble,COMPASS]{cernphprep}

\usepackage{bm}
\usepackage{color}

\usepackage{amssymb}
\usepackage{amsmath}
\usepackage{amsbsy}
\usepackage{times}
\usepackage{epsfig}
\usepackage{colordvi}
\usepackage{graphicx}
\usepackage{wrapfig,rotating}
\usepackage[numbers, square, comma, sort&compress]{natbib}
\usepackage{hyperref} 
\usepackage{multicol}
\usepackage{booktabs}
\usepackage{multirow}
\usepackage{lineno}
\RequirePackage[T1]{fontenc}

\newsavebox\myboxA
\newsavebox\myboxB
\newlength\mylenA

\newcommand{\eg}{{\it e.g.\ }}

\def\picwidth{.55}

\makeatletter

\DeclareSymbolFont{letters}     {OML}{cmm}{m}{it}
\DeclareSymbolFont{symbols}     {OMS}{cmsy}{m}{n}
\DeclareSymbolFont{largesymbols}{OMX}{cmex}{m}{n}
\begin{document}
\begin{titlepage}
\PHnumber{2014--109}
\PHdate{21 May 2014}

\title{Measurement of the charged-pion polarisability}

\Collaboration{The COMPASS Collaboration}
\ShortAuthor{The COMPASS Collaboration}

\begin{abstract}
The COMPASS collaboration at CERN has investigated pion Compton scattering,
$\pi^-\gamma\rightarrow \pi^-\gamma$, at centre-of-mass energy below 3.5 pion
masses.  The process is embedded in the reaction
$\pi^-\mathrm{Ni}\rightarrow\pi^-\gamma\;\mathrm{Ni}$, which is initiated by
190\,GeV pions impinging on a nickel target.  The exchange of quasi-real photons
is selected by isolating the sharp Coulomb peak observed at smallest momentum
transfers, $Q^2<0.0015$\,(GeV/$c$)$^2$. From a sample of 63\,000 events the pion
electric polarisability is determined to be
$\alpha_\pi\ =\ (\,2.0\ \pm\ 0.6_{\mbox{\scriptsize
    stat}}\ \pm\ 0.7_{\mbox{\scriptsize syst}}\,) \times 10^{-4}\,\mbox{fm}^3$
under the assumption $\alpha_\pi=-\beta_\pi$, which relates the electric and
magnetic dipole polarisabilities. It is the most precise measurement of this
fundamental low-energy parameter of strong interaction, that has been addressed
since long by various methods with conflicting outcomes.  While this result is
in tension with previous dedicated measurements, it is found in agreement with
the expectation from chiral perturbation theory. An additional measurement
replacing pions by muons, for which the cross-section behavior is unambigiously
known, was performed for an independent estimate of the systematic uncertainty.
\end{abstract}
\vfill
\Submitted{(to be submitted to Physical Review Letters)}
\end{titlepage}

{\pagestyle{empty}
%
%

\section*{The COMPASS Collaboration}
\label{app:collab}
\renewcommand\labelenumi{\textsuperscript{\theenumi}~}
\renewcommand\theenumi{\arabic{enumi}}
\begin{flushleft}
C.~Adolph\Irefn{erlangen},
R.~Akhunzyanov\Irefn{dubna}, 
M.G.~Alexeev\Irefn{turin_u},
G.D.~Alexeev\Irefn{dubna}, 
A.~Amoroso\Irefnn{turin_u}{turin_i},
V.~Andrieux\Irefn{saclay},
V.~Anosov\Irefn{dubna}, 
A.~Austregesilo\Irefnn{cern}{munichtu},
B.~Bade{\l}ek\Irefn{warsawu},
F.~Balestra\Irefnn{turin_u}{turin_i}
% ...
\end{flushleft}

%
%

\begin{Authlist}
\item \Idef{bielefeld}{Universit\"at Bielefeld, Fakult\"at f\"ur Physik, 33501 Bielefeld, Germany\Arefs{f}}
\item \Idef{bochum}{Universit\"at Bochum, Institut f\"ur Experimentalphysik, 44780 Bochum, Germany\Arefs{f}\Arefs{ll}}
\item \Idef{bonniskp}{Universit\"at Bonn, Helmholtz-Institut f\"ur  Strahlen- und Kernphysik, 53115 Bonn, Germany\Arefs{f}}
\item \Idef{bonnpi}{Universit\"at Bonn, Physikalisches Institut, 53115 Bonn, Germany\Arefs{f}}
\item \Idef{brno}{Institute of Scientific Instruments, AS CR, 61264 Brno, Czech Republic\Arefs{g}}
\item \Idef{calcutta}{Matrivani Institute of Experimental Research \& Education, Calcutta-700 030, India\Arefs{h}}
\item \Idef{dubna}{Joint Institute for Nuclear Research, 141980 Dubna, Moscow region, Russia\Arefs{i}}
\item \Idef{erlangen}{Universit\"at Erlangen--N\"urnberg, Physikalisches Institut, 91054 Erlangen, Germany\Arefs{f}}
\item \Idef{freiburg}{Universit\"at Freiburg, Physikalisches Institut, 79104 Freiburg, Germany\Arefs{f}\Arefs{ll}}
\item \Idef{cern}{CERN, 1211 Geneva 23, Switzerland}
\item \Idef{liberec}{Technical University in Liberec, 46117 Liberec, Czech Republic\Arefs{g}}
\item \Idef{lisbon}{LIP, 1000-149 Lisbon, Portugal\Arefs{j}}
\item \Idef{mainz}{Universit\"at Mainz, Institut f\"ur Kernphysik, 55099 Mainz, Germany\Arefs{f}}
\item \Idef{miyazaki}{University of Miyazaki, Miyazaki 889-2192, Japan\Arefs{k}}
\item \Idef{moscowlpi}{Lebedev Physical Institute, 119991 Moscow, Russia}
\item \Idef{munichlmu}{Ludwig-Maximilians-Universit\"at M\"unchen, Department f\"ur Physik, 80799 Munich, Germany\Arefs{f}\Arefs{l}}
\item \Idef{munichtu}{Technische Universit\"at M\"unchen, Physik Department, 85748 Garching, Germany\Arefs{f}\Arefs{l}}
\item \Idef{nagoya}{Nagoya University, 464 Nagoya, Japan\Arefs{k}}
\item \Idef{praguecu}{Charles University in Prague, Faculty of Mathematics and Physics, 18000 Prague, Czech Republic\Arefs{g}}
\item \Idef{praguectu}{Czech Technical University in Prague, 16636 Prague, Czech Republic\Arefs{g}}
\item \Idef{protvino}{State Scientific Center Institute for High Energy Physics of National Research Center `Kurchatov Institute', 142281 Protvino, Russia}
\item \Idef{saclay}{CEA IRFU/SPhN Saclay, 91191 Gif-sur-Yvette, France\Arefs{ll}}
\item \Idef{telaviv}{Tel Aviv University, School of Physics and Astronomy, 69978 Tel Aviv, Israel\Arefs{m}}
\item \Idef{triest_u}{University of Trieste, Department of Physics, 34127 Trieste, Italy}
\item \Idef{triest_i}{Trieste Section of INFN, 34127 Trieste, Italy}
\item \Idef{triest_ictp}{Abdus Salam ICTP, 34151 Trieste, Italy}
\item \Idef{turin_u}{University of Turin, Department of Physics, 10125 Turin, Italy}
\item \Idef{turin_p}{University of Eastern Piedmont, 15100 Alessandria, Italy}
\item \Idef{turin_i}{Torino Section of INFN, 10125 Turin, Italy}
\item \Idef{warsaw}{National Centre for Nuclear Research, 00-681 Warsaw, Poland\Arefs{n} }
\item \Idef{warsawu}{University of Warsaw, Faculty of Physics, 00-681 Warsaw, Poland\Arefs{n} }
\item \Idef{warsawtu}{Warsaw University of Technology, Institute of Radioelectronics, 00-665 Warsaw, Poland\Arefs{n} }
\item \Idef{yamagata}{Yamagata University, Yamagata, 992-8510 Japan\Arefs{k} }
\end{Authlist}
%
%
\vspace*{-\baselineskip}\renewcommand\theenumi{\alph{enumi}}
\begin{Authlist}
\item \Adef{a}{Also at Instituto Superior T\'ecnico, Universidade de Lisboa, Lisbon, Portugal}
\item \Adef{aa}{Also at Department of Physics, Pusan National University, Busan 609-735, Republic of Korea and at Physics Department, Brookhaven National Laboratory, Upton, NY 11973, U.S.A. }
\item \Adef{bb}{Supported by the DFG Research Training Group Programme 1102  ``Physics at Hadron Accelerators''}
\item \Adef{b}{Also at Chubu University, Kasugai, Aichi, 487-8501 Japan\Arefs{k}}
\item \Adef{c}{Also at KEK, 1-1 Oho, Tsukuba, Ibaraki, 305-0801 Japan}
\item \Adef{c1c}{Present address: Universit\"at Bonn, Helmholtz-Institut f\"ur Strahlen- und Kernphysik, 53115 Bonn, Germany}
\item \Adef{cc}{Also at Moscow Institute of Physics and Technology, Moscow Region, 141700, Russia}
\item \Adef{x}{present address: RWTH Aachen University, III. Physikalisches Institut, 52056 Aachen, Germany}
\item \Adef{f}{Supported by the German Bundesministerium f\"ur Bildung und Forschung}
\item \Adef{g}{Supported by Czech Republic MEYS Grants ME492 and LA242}
\item \Adef{h}{Supported by SAIL (CSR), Govt.\ of India}
\item \Adef{i}{Supported by CERN-RFBR Grants 08-02-91009 and 12-02-91500}
\item \Adef{j}{\raggedright Supported by the Portuguese FCT - Funda\c{c}\~{a}o para a Ci\^{e}ncia e Tecnologia, COMPETE and QREN, Grants CERN/FP/109323/2009, CERN/FP/116376/2010 and CERN/FP/123600/2011}
\item \Adef{k}{Supported by the MEXT and the JSPS under the Grants No.18002006, No.20540299 and No.18540281; Daiko Foundation and Yamada Foundation}
\item \Adef{l}{Supported by the DFG cluster of excellence `Origin and Structure of the Universe' (www.universe-cluster.de)}
\item \Adef{ll}{Supported by EU FP7 (HadronPhysics3, Grant Agreement number 283286)}
\item \Adef{m}{Supported by the Israel Science Foundation, founded by the Israel Academy of Sciences and Humanities}
\item \Adef{n}{Supported by the Polish NCN Grant DEC-2011/01/M/ST2/02350}
\item [{\makebox[2mm][l]{\textsuperscript{*}}}] Deceased
\end{Authlist}

\newpage
The electric and magnetic polarisabilities of an extended object describe its
rigidity against deformation by external electric and magnetic fields,
respectively.  For a strongly interacting particle, the polarisabilities are of
special interest as they are related to the inner forces determining the
substructure and thus provide valuable information about quantum chromodynamics
(QCD) at low energy.  
The pion is of specific interest in that regard, as it represents the lightest QCD bound state 
and its polarisability, once experimentally determined, imposes stringent constraints on theory as discussed below.

For the proton, the polarisability is measured directly
via Compton scattering on a hydrogen target.  In contrast, for charged pions the
experimental situation is more difficult since they are not available as fixed
target.  Although different techniques exist, all previous measurements are
affected by large experimental and theoretical uncertainties, see \eg
Refs. \cite{Antipov:1982kz, Ahrens:2004mg, Boyer:1990}. Groundbreaking work at
Serpukhov~\cite{Antipov:1982kz} employed the same Primakoff technique
\cite{Primakoff:1951pj} as used in this Letter, however low statistics made it
difficult at that time to evaluate the systematic uncertainty.

The electric and magnetic dipole polarisabilities $\alpha_\pi$ and $\beta_\pi$
appear at the level of the pion Compton cross section $\sigma_{\pi \gamma}$ for
the reaction $\pi^-\gamma\rightarrow\pi^-\gamma$ in the correction to the Born
cross section for the point-like particle at linear
order~\cite{Petrunkin:1964,Drechsel:1994kh} as
\begin{equation}
\label{eq:1}
\frac{d\sigma_{\pi\gamma}}{d\Omega} =
\left(\frac{d\sigma_{\pi\gamma}}{d\Omega}\right)_\mathrm{Born} -\frac{\alpha\,
  m_\pi^3(s-m_\pi^2)^2}{4s^2\,(s\, z_+ + m_\pi^2\, z_-)}
\left(z_-^2(\alpha_\pi-\beta_\pi)
+\frac{s^2}{m_\pi^4}z_+^2(\alpha_\pi+\beta_\pi)\right).
\end{equation}
Here $\alpha\approx 1/137.04$ is the fine structure constant,
$z_\pm=1\pm\cos\theta_\mathrm{cm}$ with $\theta_\mathrm{cm}$ being the $\pi\gamma$ scattering
angle, $s$ is the squared total energy in the center-of-mass reference frame,
and $m_\pi$ is the rest mass of the charged pion. Higher-order contributions 
can be parameterised by further multipole polarisabilities, which are neglected in this analysis.

For hadronic interactions at low energy, QCD can be formulated in terms of an
effective field theory that results from the systematic treatment of chiral
symmetry and its breaking pattern, which is called chiral perturbation theory
(ChPT).  In this approach, the pions ($\pi^+,\pi^0,\pi^-$) are identified with
the Goldstone bosons associated with spontaneous chiral symmetry breaking.
Properties and interactions of pions hence provide the most rigorous test
whether ChPT is the correct low-energy representation of QCD.  The predictions
for the dynamics of low-energy $\pi\pi$ scattering were confirmed in various
experiments, see \mbox{\eg Ref. \cite{Adolph:2011it}}.  However, in the case of
$\pi\gamma$ scattering the ``Serpukhov \mbox{value'' $\alpha_{\pi}=(6.8\pm1.8)
  \times 10^{-4}$\,fm$^3$\cite{Antipov:1982kz}} for the pion polarisability
deviates from the ChPT prediction $\alpha_{\pi}=(2.9\pm0.5) \times
10^{-4}$\,fm$^3$ \cite{Gasser:2006qa}.  This observation, which was confirmed in
radiative pion photoproduction at MAMI~\cite{Ahrens:2004mg}, remained
unexplained for more than two decades.

In pion-nucleus reactions, photon exchange becomes important at very low
momentum transfer and competes with strong interaction processes.  The
$\pi$-nucleus cross section can be connected to the $\pi\gamma$ cross section
using the equivalent-photon approximation (EPA)~\cite{Pomeranshuk:1961}:
\begin{equation}
  \label{eq:primbasic}
        {\frac{d\sigma_{\mbox{\tiny (A,Z)}}^{\mbox{\tiny EPA}}}
          {ds\,dQ^2\,d\Phi_{n}}} = {\frac{Z^{2}\alpha}{\pi
            (s-m_{\pi}^{2})}}\ F^2(Q^2)\ \frac{Q^2-Q_{\min}^2}{Q^{4}}
        \ \frac{d\sigma_{\pi\gamma\rightarrow X}}{d\Phi_{n}} .
\end{equation}
Here, the cross section for the process $\pi^-$(A, Z)$\rightarrow X^-$\,(A, Z)
is factorized into the quasi-real photon density provided by the nucleus of
charge $Z$, and $\sigma_{\pi\gamma\rightarrow X}$ denotes the cross section for
the embedded $\pi^-\gamma\rightarrow X^-$ reaction of a pion and a real
photon. The function $F(Q^2)$ is the electromagnetic form factor of the nucleus
and $d\Phi_{n}$ is the $n$-particle phase-space element of the final-state
system $X^-$.  The minimum value of the negative 4-momentum transfer squared,
$Q^2=-(p_{\mbox{\tiny beam}}^{\mu}-p_X^{\mu})^{2}$, is
\mbox{$Q^2_{\min}=(s-m_{\pi}^2)^2/(4E_{\mbox{\tiny beam}}^{2})$} for a given
final-state mass $m_X=\sqrt{s}$, with typical values $Q^2_{\min}=(1$\,MeV/$c$)$^2$.
In the analysis presented in this Letter, the
observed final state is $\pi^-\gamma$, and the investigated cross section
$\sigma_{\pi\gamma\rightarrow X}$ is $\sigma_{\pi\gamma}$ as introduced along
with Eq.~(\ref{eq:1}) with $s=(p_\pi^{\mu}+p_\gamma^{\mu})^2$ being determined
by the 4-vectors of the two outgoing particles.  The same experimental technique
has been employed previously at COMPASS for the $\pi^-\pi^-\pi^+$ final state
\cite{Adolph:2011it}.

\begin{figure}
 \begin{center} 
   \includegraphics[width=220px]{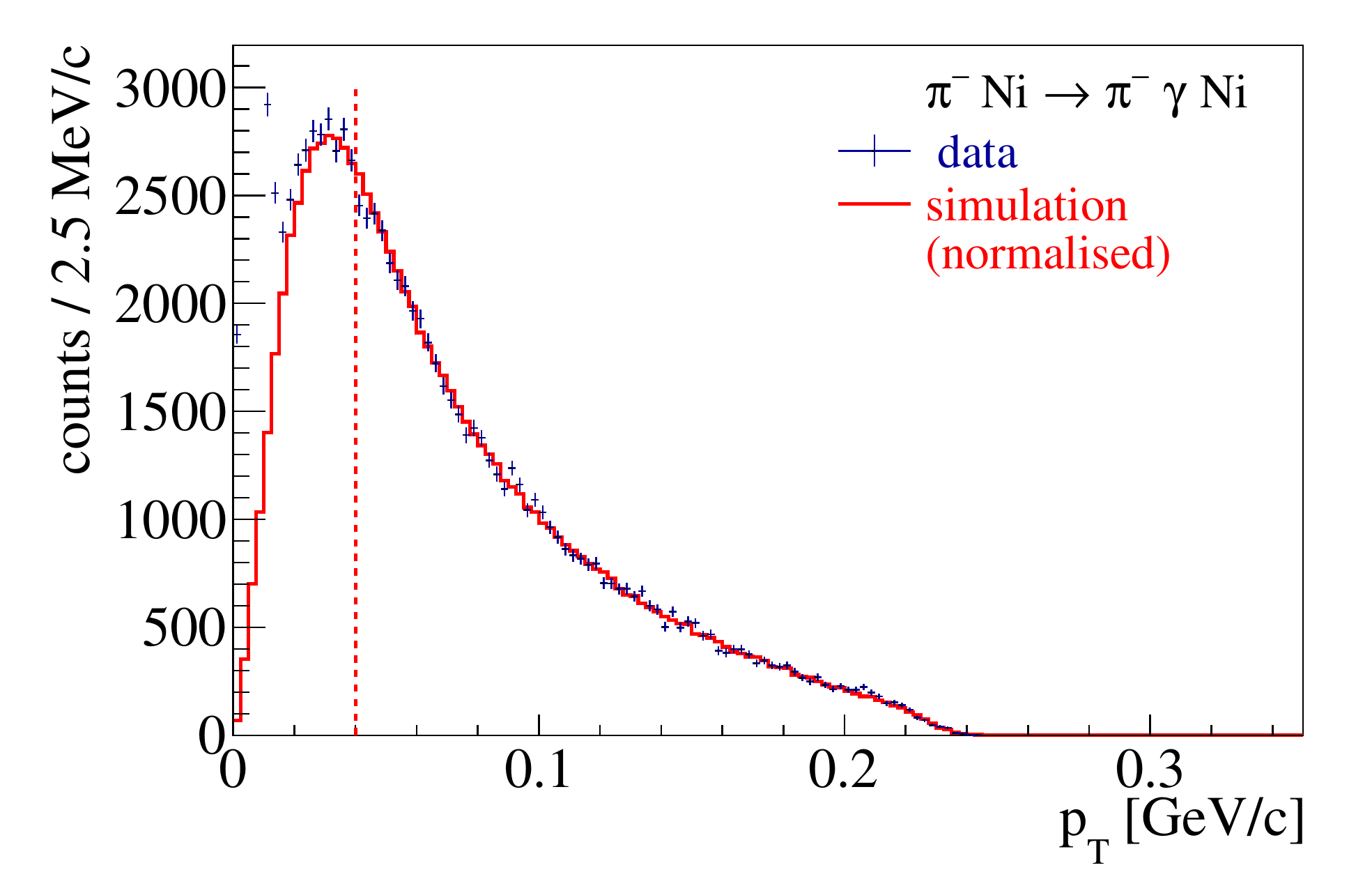}
   \includegraphics[width=220px]{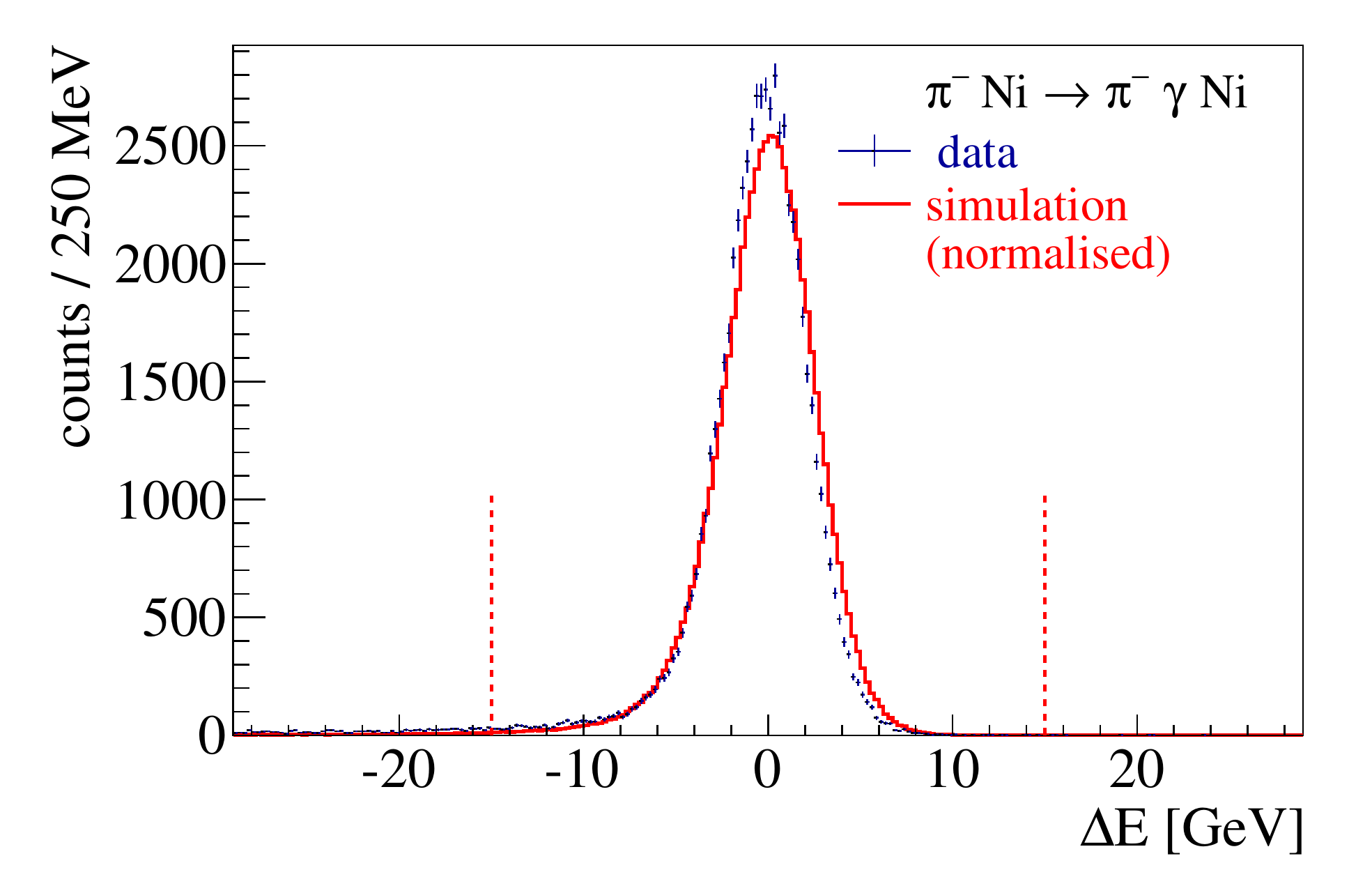}
   (a)\hspace{0.4\textwidth}(b)$\quad$
   \\ \includegraphics[width=220px]{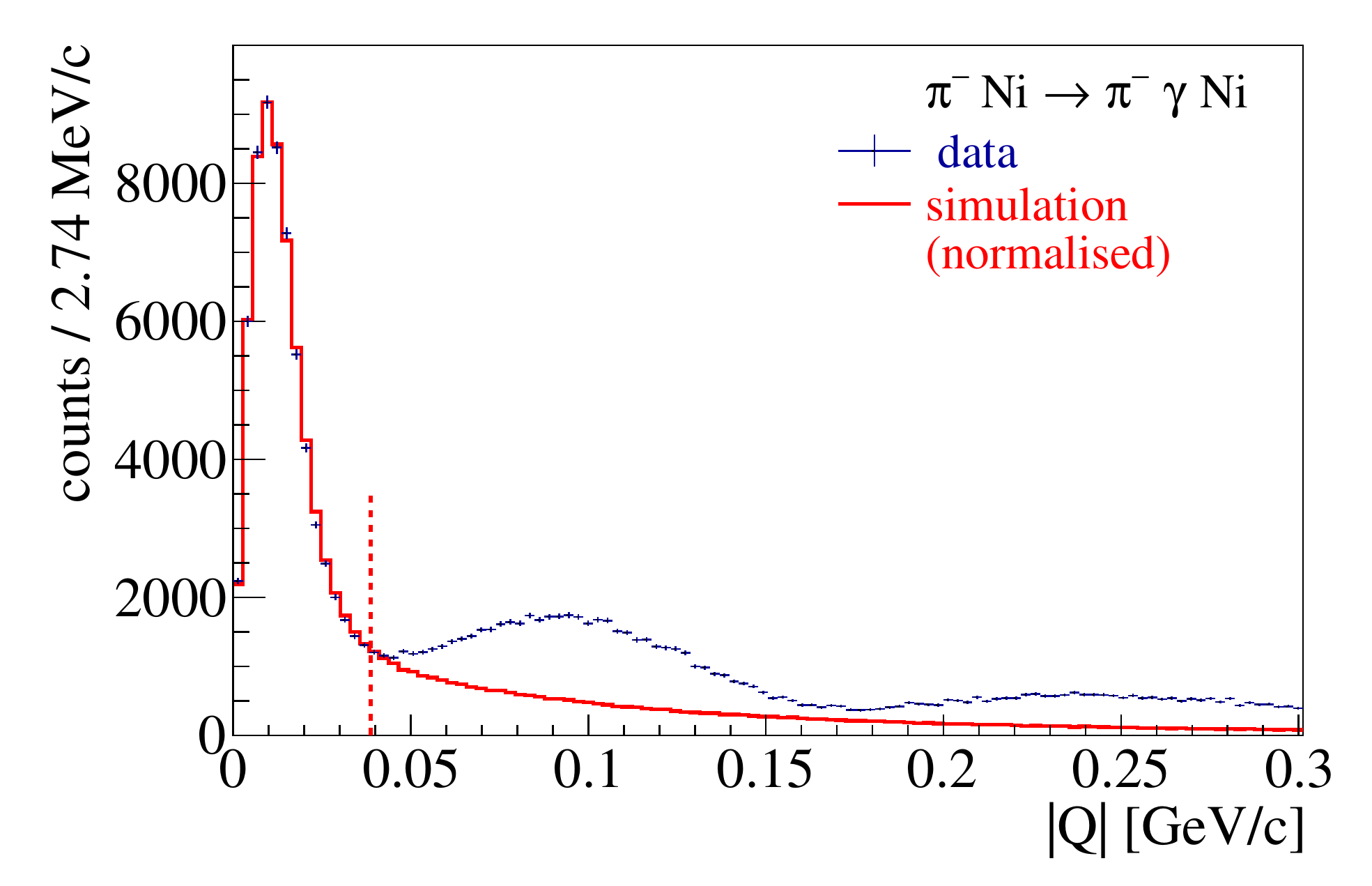}
   \includegraphics[width=220px]{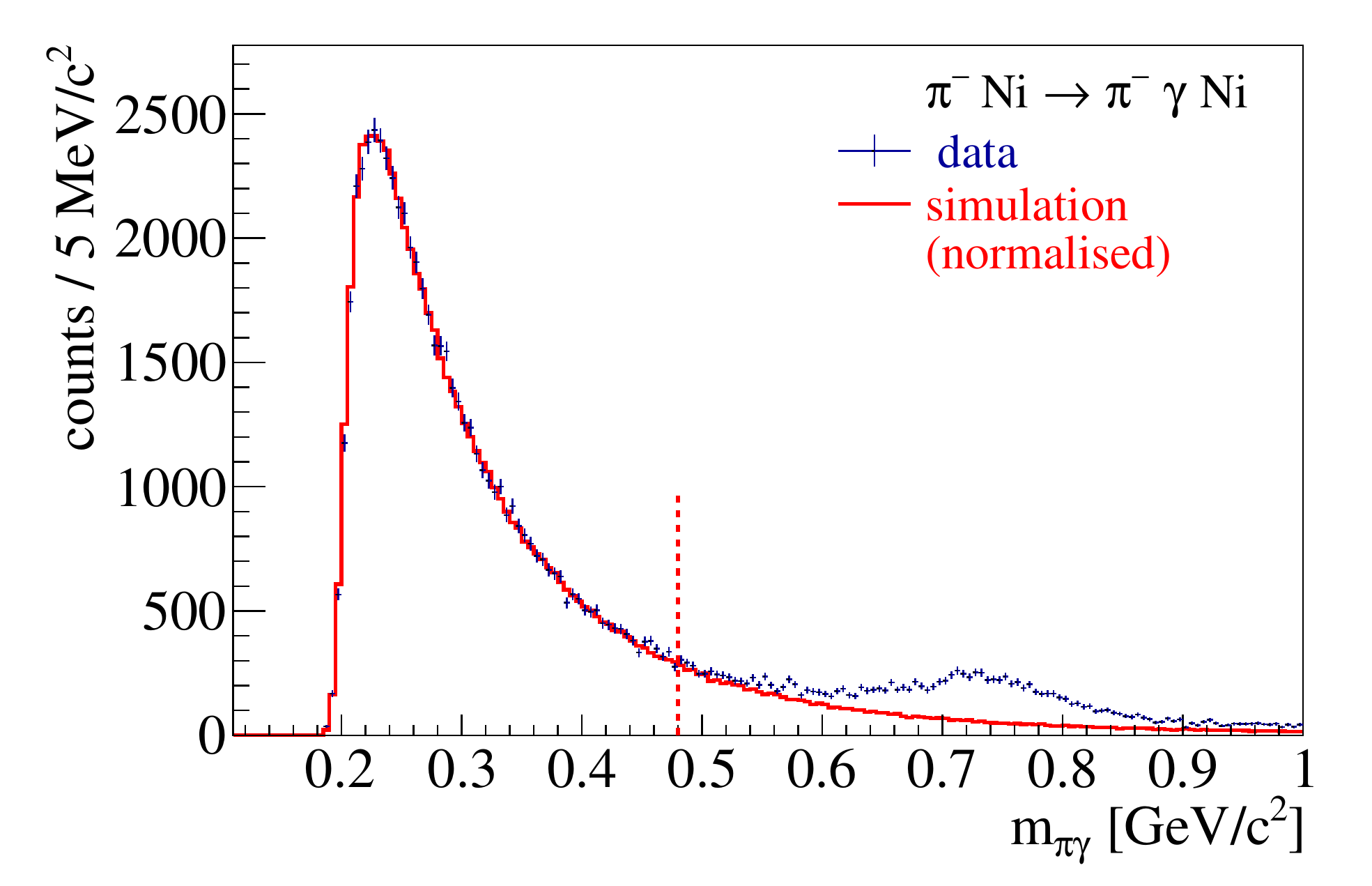}
   (c)\hspace{0.4\textwidth}(d)$\quad$
 \end{center}
  \caption{\label{fig:pionkin} Comparison of the measured (black points with
    error bars) and simulated (red histograms) kinematic distributions for
    measurements with pion beam: (a) transverse momentum $p_{T}$ of the
    scattered pion; (b) energy balance $\Delta E$; (c) $|Q|$ distribution,
    featuring for the real data at higher values the contribution from strong
    interaction, which is not contained in the simulation; (d) invariant mass of
    the $\pi\gamma$ system. The dotted lines indicate the cuts as explained in
    the text.} 
\end{figure}

The COMPASS experiment \cite{Abbon:2007pq} is situated at the M2 beam line of
the CERN Super Proton Synchrotron. For this measurement, negative muons or
hadrons of 190\,GeV/$c$ were used, which were impinging on a 4\,mm thick nickel
target. The hadronic components of the hadron beam at the target position are
\mbox{96.8\% $\pi^-$}, \mbox{2.4\% $K^-$} and \mbox{0.8\% $\bar{p}$}. The hadron
beam also contains about 1\% of muons and a small amount of electrons. The pions
are identified with a Cherenkov counter located in the beam line at the entrance
to the experimental area.  The large-acceptance high-precision spectrometer is
well suited for investigations of high-energy reactions at low to intermediate
momentum transfer to the target nucleus.  Outgoing charged particles are
detected by the tracking system and their momenta are determined using two large
aperture magnets. Tracks crossing more than 15 radiation lengths equivalent
thickness of material are treated as muons. The small-angle electromagnetic
calorimeter ECAL2 detects photons up to scattering angles of about 40\,mrad.

The data presented in this Letter were recorded in the year 2009 using
alternatively either hadron or muon beams. The trigger logic selects events with
an energy deposit of more than 70\,GeV in the central part of ECAL2 in
coincidence with an incoming beam particle.  In the data analysis, exactly one
scattered, negatively charged particle, which is assumed to be a pion, is
required to form with the incoming pion a vertex that is consistent with an
interaction in the target volume. Exactly one cluster in ECAL2 with an energy
above 2\,GeV, which is not attributed to a produced charged particle, is
required and taken as the produced photon.  In order to avoid the kinematic
region that is dominated by multiple scattering of the outgoing pion in the
target material, only events with $p_{T}>$\,40\,MeV/$c$ are accepted, as shown
in Fig.~\ref{fig:pionkin}(a).  This cut also removes contributions of the
reaction $e^-$\,Ni $\rightarrow e^-\gamma$\,Ni.  Neglecting the tiny recoil of
the target nucleus at low $Q^2$, the sum of the scattered pion energy $E_{\pi}$
and the photon energy $E_\gamma$ equals the beam energy for the exclusive
reaction $\pi^-$\,Ni$\rightarrow\pi^-\gamma$\,Ni.  The distribution of events as
a function of the energy balance $\Delta E=E_{\pi}+E_\gamma-E_{\mbox{\tiny
    beam}}$ is presented in Fig.~\ref{fig:pionkin}(b).  As the calorimetric
energy resolution is approximately constant over the range of interest and about
3\,GeV, the energy balance is required to be $|\Delta E|<$15\,GeV.  After this
selection, we assume the reaction $\pi^-$Ni$\rightarrow \pi^- \gamma$ Ni and
imposing energy conservation, we rescale the photon momentum vector such that
$E_{\gamma}=E_{\mbox{\tiny beam}}-E_{\pi}$, as the photon energy is the least
known quantity.
\begin{figure}
\begin{center}  
  \includegraphics[angle=90,width=\picwidth\linewidth]{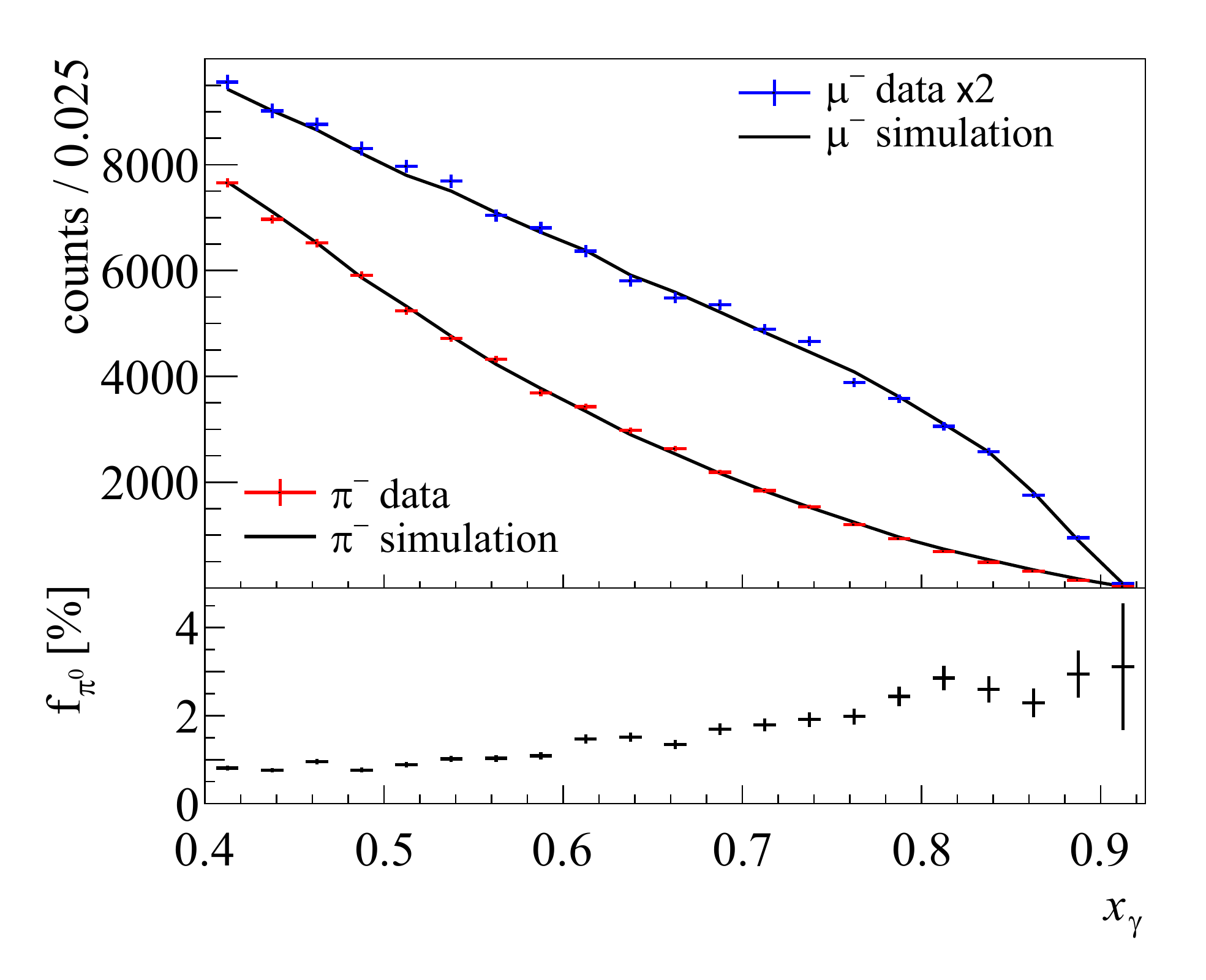}\\
  \caption{\label{fig:pionsel} The measured and simulated $x_\gamma$
    distributions for pion (lower curve) and muon (upper curve) beam. The
    statistical uncertainty of the real data points is indicated by vertical
    error bars, while the width of the symbols is set arbitrarily to one third
    of the bin width. The lines connect the simulation results for the same bin
    centers. The bottom panel shows the $\pi^0$ background fraction
    $f_{\pi^{0}}$ that was subtracted from the pion data.}
\end{center}  
\end{figure}
The distribution of events as a function of $|Q|=\sqrt{Q^2}$ is given in
Fig.~\ref{fig:pionkin}(c). The peak width of about 12 MeV/$c$ is dominated by
the experimental resolution, which is about a factor of ten larger than the true
width of the Coulomb distribution.  Events corresponding to photon exchange are
selected by requiring $Q^2<0.0015$\,(GeV/$c$)$^2$. The size of the Coulomb peak
was checked for different targets on smaller-statistics data (tungsten, silicon,
carbon), showing consistency with the approximate $\sim \!\!Z^2$
expectation. Background contributions from intermediate $\rho^{-}(770)$
production with decay into $\pi^{-}\pi^0$ are suppressed by restricting to the
mass interval \mbox{$m_{\pi\gamma}<3.5$\,$m_{\pi}\approx0.487$\,GeV/$c^2$}, as
shown in Fig.~\ref{fig:pionkin}(d).  For this analysis, we choose the region
$0.4 < x_\gamma < 0.9$, where $x_\gamma=E_\gamma/E_{\mbox{\tiny beam}}$ is the
fraction of the beam energy taken by the photon in the laboratory system. This
region is characterised by constant trigger efficiency and effective
identification of muons. The number of $\pi \gamma$ events in this region is
\mbox{63 000}.

The pion polarisability manifests itself by a modification of the differential
Compton cross section at high photon energies that correspond to large forces
exerted to the pion.  For retrieving the pion polarisability from the shape of
the measured cross section, the analysis technique as described in
Ref. \cite{Antipov:1982kz} is adopted.  This includes the assumption that
$\alpha_\pi$ is approximately equal in magnitude to the magnetic polarisability
$\beta_\pi$, but with opposite sign. In this analysis we use
$\alpha_\pi=-\beta_\pi$. The polarisability is determined from the $x_{\gamma}$
dependence of the ratio
\begin{equation}        
  R_{\pi}\ =\ \left(\frac{d\!\sigma_{\pi \gamma}}{d\!x_\gamma}\right) \left/
  \left(\frac{d\!\sigma^{0}_{\pi \gamma}}{d\!x_\gamma}\right) \right.  = 1
  \ -\ \frac{3}{2}\cdot \frac{m_\pi^3}{\alpha} \cdot
  \frac{x_\gamma^2}{1-x_\gamma}\ \alpha_\pi,
  \label{eq:rxg}
\end{equation}
where $\sigma_{\pi \gamma}=N/L$ refers to the measured cross section,
$d\sigma^0_{\pi \gamma}$ to the simulated cross section expected for
$\alpha_\pi=0$ (including corrections to the pure Born cross section as those
from chiral loops, as specified below), $N$ is the number of events, and $L$ is
the integrated luminosity.  The variable $x_\gamma$ is to a good approximation
related to the photon scattering angle by
$\cos\theta_\mathrm{cm}\approx1-2x_\gamma/(1-m_\pi^2/s)$, so that the selected
range in $x_\gamma$ corresponds to $-1<\cos\theta_\mathrm{cm}<0.15$, where the
sensitivity to $\alpha_\pi-\beta_\pi$ is largest, see Eq. (1).  The event
distribution in the variable $x_\gamma$ is shown in Fig.~\ref{fig:pionsel}
together with the simulated data that were generated with $\alpha_{\pi}=0$ and
scaled such that the integral is the same as for the real data, disregarding at
this point the small effect of the pion polarisability.  The requirement $\Delta
E<$15\,GeV and the observation of exactly one photon in ECAL2 do not completely
eliminate the background from $\pi^0$ mesons produced in electromagnetic and
strong interactions, $\pi^-$ Ni $\rightarrow \pi^-\pi^0$ X, where in the
considered low $Q^2$ region X is predominantly a Ni nucleus in its ground-state,
but in principle nuclear excitation or breakup is also included. The probability
to misidentify such $\pi^-\pi^0$ events as $\pi^-\gamma$ events due to missing
or overlapping photons is estimated from a pure sample of beam kaon decays,
$K^-\rightarrow \pi^- \pi^0$ , and the observation of corresponding (in this
case unphysical) $\pi^-\gamma$ final states. The same probability is assumed for
misidentifying $\pi^-\pi^0$ as $\pi^-\gamma$ for the studied $\pi^-$Ni reactions
in each $x_{\gamma}$ bin, and the fraction $f_{\pi^0}$ of background caused by
$\pi^{0}$ events is presented in the bottom panel of Fig.~\ref{fig:pionsel}. 
a function of $x_{\gamma}$.  The simulated cross section $d\!\sigma^{0}_{\pi
  \gamma}/d\!x_\gamma$ contains besides the Born term the following corrections:
i) radiative corrections \cite{Kaiser:2008jm}; ii) chiral loop corrections
\cite{Kaiser:2008ss}; iii) corrections for the electromagnetic form factor of
the nickel nucleus, which is approximated for simplicity by the equivalent
sharp-radius formula $F(Q^2)=j_1(r q)$ with $r= 5.0\,\mathrm{fm}$, where $q$ is
the modulus of the 3-momentum transfer to the nucleus.  More precise form-factor
parameterisations were checked with no visible influence on the results.  These
corrections influence the $x_\gamma$ spectrum such that the extracted
polarisibility is increased by $0.6 \times 10^{-4}$\,fm$^3$ after they are
applied.  The ratio of the measured differential cross section
$d\sigma_{\pi\gamma}/dx_{\gamma}$ to the expected cross section for a point-like
spin-0 particle taken from the simulation is shown in the top panel of
Fig. \ref{fig:pionres}.  The fit of the ratio $R_{\pi}$ by Eq.\,(\ref{eq:rxg})
in the range $0.4< x_{\gamma}< 0.9$, using the integrated luminosity $L$ as
additional free parameter, yields the pion polarisability:
$\alpha_\pi\ =\ (2.0\ \pm\ 0.6_{\mbox{\scriptsize stat}}) \times
10^{-4}\,\mbox{fm}^3$.

The systematic uncertainty of the measurement, as summarized in
Table~\ref{tab:systematics2}, accounts for: i) uncertainty of the determination
of the tracking detector efficiency for the simulation; ii) uncertainty related
to the neglect of Coulomb corrections \cite{Faldt} and of corrections for
nuclear charge screening by atomic electrons and for multiple-photon exchange
\cite{Andreev/Bugaev}; iii) statistical uncertainty of the $\pi^{0}$ background
subtraction; \mbox{iv) effect of} the uncertainty on the estimate of strong
interaction background and its interference with the Coulomb contribution; v)
contribution from the elastic pion-electron scattering process; vi) contribution
from the $\mu^-$ Ni $\rightarrow \mu^-\gamma$ Ni reaction, where the scattered
muon was misidentified as pion.  The total systematic uncertainty is obtained by
adding these six contributions in quadrature. The final result on the pion
polarisability is:
\begin{equation}
\alpha_\pi\ =\ (2.0\ \pm\ 0.6_{\mbox{\scriptsize stat}}
\ \pm\ 0.7_{\mbox{\scriptsize syst}}) \times 10^{-4}\,\mbox{fm}^3.
\label{eq:pol2009stat}
\end{equation}
A measurement with the pion beam replaced by a muon beam of the same momentum
was performed in order to validate the result obtained for the pion cross
section $d\!\sigma_{\pi\gamma}/d\!x_\gamma$.  The same selection criteria as
used for the pion sample are applied adapting the cut $m_{\mu\gamma}<3.5\,m_{\mu}$.
The simulation for the muon measurement contains the corresponding
radiative \cite{Kaiser/Arbuzov} and form factor corrections. Taking into account
the different behavior of the cross section for a point-like spin-$\frac{1}{2}$
particle, no deviation from the QED prediction is expected for the muon.  Using
the measurement with the muon beam, the ``false polarisability'' is determined
from the $x_{\gamma}$ dependence of the ratio $R_{\mu}$, that is defined
analogously to Eq.~(\ref{eq:rxg}).  It is found to be compatible with zero
within statistical uncertainties, $(0.5\pm0.5_{\mbox{\scriptsize
    stat}}) \times 10^{-4}$\,fm$^3$, as shown in the bottom panel of
Fig. \ref{fig:pionres}.
\begin{table}
\begin{center}
  \caption{Estimated systematic uncertainties at 68\,\% confidence level.}
  \label{tab:systematics2}
 \begin{tabular}{lr}
    \toprule \multirow{2}{*}{Source of uncertainty} & Estimated magnitude\\ &
    $\quad[10^{-4}\,\text{fm}\rule{0pt}{7pt}^3]$\\ \midrule Determination of
    tracking detector efficiency & 0.5\\ Treatment of radiative corrections &
    0.3\\ Subtraction of $\pi^{0}$ background & 0.2\\ Strong interaction
    background & 0.2\\ Pion-electron elastic scattering & 0.2\\ Contribution of
    muons in the beam & 0.05 \\ \midrule \midrule Quadratic sum &
    0.7\\ \bottomrule
  \end{tabular}
\end{center}
\end{table}

\begin{figure}
\begin{center}  
  \includegraphics[width=\picwidth\linewidth]{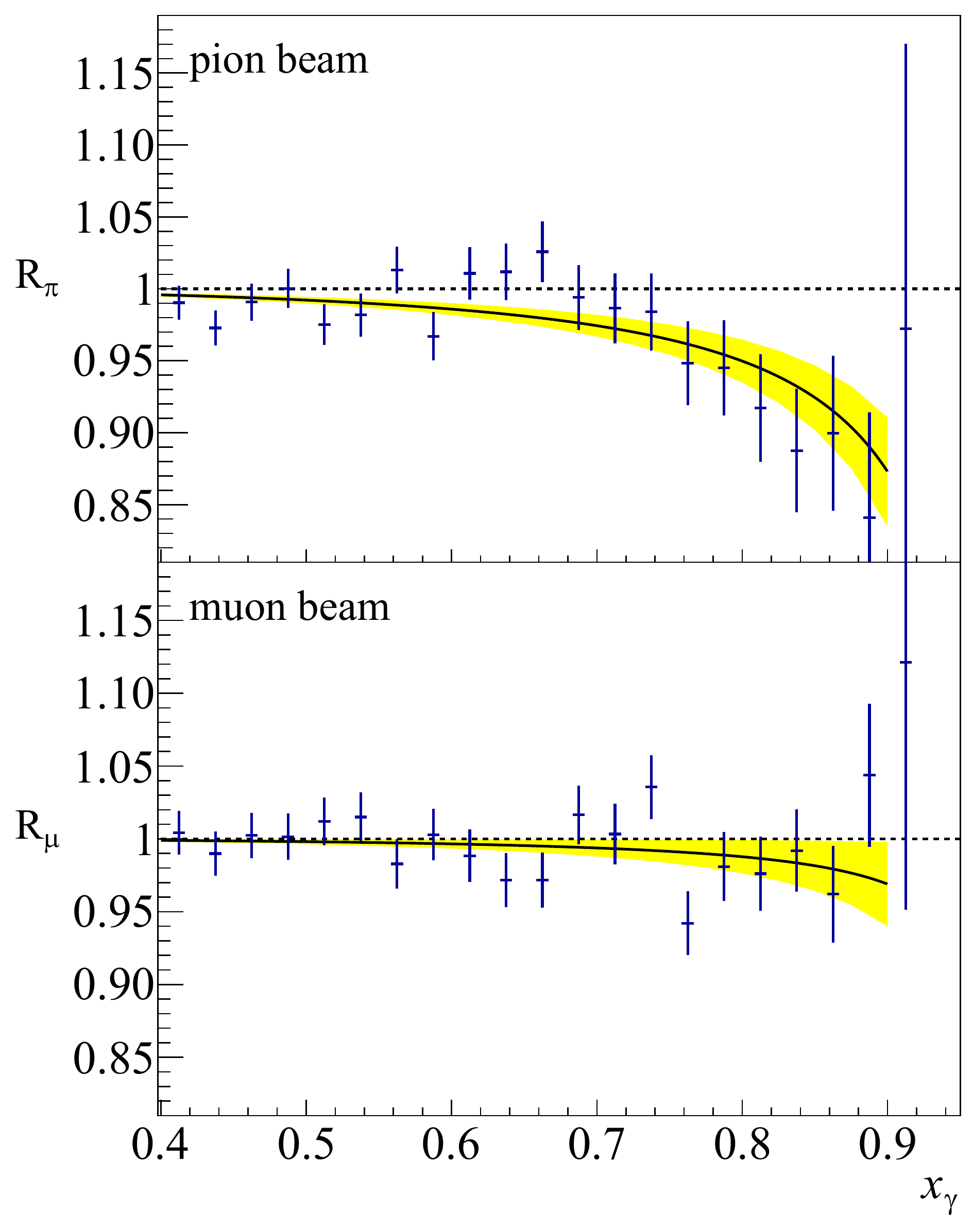}\\
  \caption{\label{fig:pionres} The $x_\gamma$ dependence of the ratio of the
    measured differential cross section $d\sigma/dx_{\gamma}$ over the expected
    cross section for point-like particles. Top (bottom) panel: measurement with
    pion (muon) beam. The respective ratios contain the corrections described in
    the text. The bands denote the respective statistical uncertainties of the
    fit results shown by the solid lines. Error bars denote statistical
    uncertainties. The quality of the fits can be characterized by the values
		$\chi_{\pi}^2/\text{NDF}=22.0/18$ and $\chi_{\mu}^2/\text{NDF}=19.6/18$, respectively.}
\end{center}  
\end{figure}
Possible contributions from higher-order polarisabilities beyond
Eq.~(\ref{eq:1}), were studied by investigating the sensitivity of the result on
the upper limit of $m_{\pi\gamma}$. No significant effect was found when varying
this limit between 0.40 GeV/c$^2$ and 0.57 GeV/c$^2$. Furthermore, the
functional behavior of our model, including the chiral-loop corrections, was
compared to the approach using dispersion relations \cite{Pasquini}, and very
good agreement was found in the mass range up to 4m$_{\pi}$. The respective
cross sections do not differ by more than 2 permille, which corresponds to less
than 15\% of the given systematic uncertainty estimate for the polarisability
value.

In conclusion, we have determined the pion polarisability from pion Compton
scattering embedded in the $\pi^-$\,Ni$\rightarrow\pi^-\gamma$\,Ni process at
small momentum transfer, $Q^2<0.0015$\,(GeV/$c$)$^2$. The measurement using a
muon beam has revealed no systematic bias of our method. We find the size of the
pion polarisability at significant variance with previous experiments and
compatible with the expectation from ChPT. This result constitutes important
progress towards resolving one of the long-standing issues in low energy QCD.

We gratefully acknowledge the support of the CERN management and staff as well
as the skills and efforts of the technicians of the collaborating institutions.
This work is supported by MEYS (Czech Republic); ``HadronPhysics2'' Integrating
Activity in FP7 (European Union); CEA, P2I and ANR (France); BMBF, DFG cluster
of excellence ``Origin and Structure of the Universe'', the computing facilities
of the Computational Center for Particle and Astrophysics (C2PAP), IAS-TUM and
Humboldt foundation (Germany); SAIL (CSR) (India); ISF (Israel); INFN (Italy);
MEXT, JSPS, Daiko and Yamada Foundations (Japan); NRF (Rep. of Korea); NCN
(Poland); FCT (Portugal) and CERN-RFBR (Russia).

\end{document}